\documentclass[11pt,twoside]{article}
\usepackage{./asp2014}
\usepackage{epstopdf}

\aspSuppressVolSlug
\resetcounters

\begin{document}

\title{Point Source Detection Software in the SKA Era}
\author{Siamak~Dehghan,$^1$ Melanie~Johnston-Hollitt,$^1$ and Christopher~Hollitt$^2$
\affil{$^1$School of Chemical and Physical Sciences, Victoria University of Wellington, P.O. Box 600, Wellington 6140, New Zealand; \email{siamak.dehghan@vuw.ac.nz}, \email{melanie.johnston-hollitt@vuw.ac.nz}}
\affil{$^2$School of Engineering and Computer Science, Victoria University of Wellington, P.O. Box 600, Wellington 6140, New Zealand; \email{christopher.hollitt@ecs.vuw.ac.nz}}}

\paperauthor{Siamak~Dehghan}{siamak.dehghan@vuw.ac.nz}{}{Victoria University of Wellington}{School of Chemical and Physical Sciences}{Wellington}{Wellington}{6140}{New Zealand}
\paperauthor{Melanie~Johnston-Hollitt}{melanie.johnston-hollitt@vuw.ac.nz}{}{Victoria University of Wellington}{School of Chemical and Physical Sciences}{Wellington}{Wellington}{6140}{New Zealand}
\paperauthor{Christopher~Hollitt}{christopher.hollitt@ecs.vuw.ac.nz}{}{Victoria University of Wellington}{School of Engineering and Computer Science}{Wellington}{Wellington}{6140}{New Zealand}

\begin{abstract}

The generation of a sky model for calibration of Square Kilometre Array observations requires a fast method of automatic point source detection and characterisation. In recent years, point source detection in two-dimensional images has been implemented by using several thresholding approaches. In the first phase of the SKA we will need a fast implementation capable of dealing with very large images ($80,000 \times 80,000$ pixels). While the underlying algorithms scale suitably with image size, the present implementations do not. We make some comments on the pertinent trade-offs for scaling these implementations to SKA-levels.

\end{abstract}

\articlefigure[width=10.9cm]{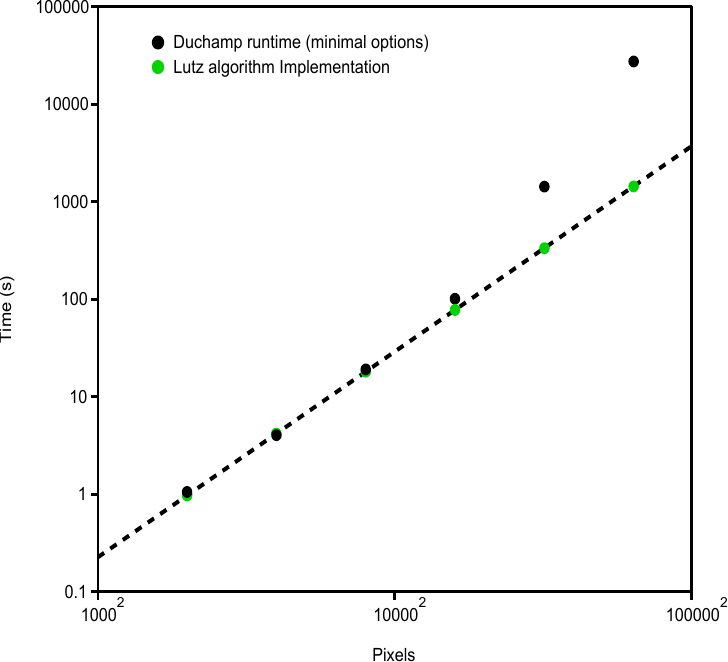}{P026_f1}{Test of Duchamp scalability. Time-complexity of Duchamp and pure Lutz algorithm implementation is shown by black and green circles, respectively. Duchamp processing time represents a deviation from linearity when analysing larger (above $16,000 \times 16,000$ pixels) images, while the running time of the underlying algorithm, Lutz, stays linear.}

\section{Introduction}

The advent of precursor radio interferometers and pathfinder technologies for the Square Kilometre Array (SKA), such as the Murchison Widefield Array (MWA) and the Low Frequency Array (LOFAR), has revealed a necessity for fast, optimal and reliable point source detection software. This is a process that has a crucial role in generation of the local sky model, and therefore, calibration of the radio observations.

Most of the current source finding packages in radio astronomy incorporate a two-stage process:

\begin{enumerate}
  \item Connected pixels above a given threshold will be grouped to form source islands; a flux threshold based on the background noise estimation is set to differentiate source pixels from the noise pixels. Once source pixels are identified, adjacent pixels are merged to form individual sources by utilising algorithms such as Lutz \citetext{\citealp{l1980}, implemented in Duchamp; \citealp{w12}} and Flood-fill (used in AEGEAN; \citealp{hmg12} and BLOBCAT; \citealp{hmc12}).
  \item The characteristics of these sources are obtained; source characterisation is generally achieved by fitting sources into single or multiple Gaussians, and in some cases (e.g., PyBDSM\footnote{http://www.astron.nl/citt/pybdsm/}), alternative basis sets, such as shapelet or wavelet decomposition are used as methods of describing sources.
\end{enumerate}

\articlefigure{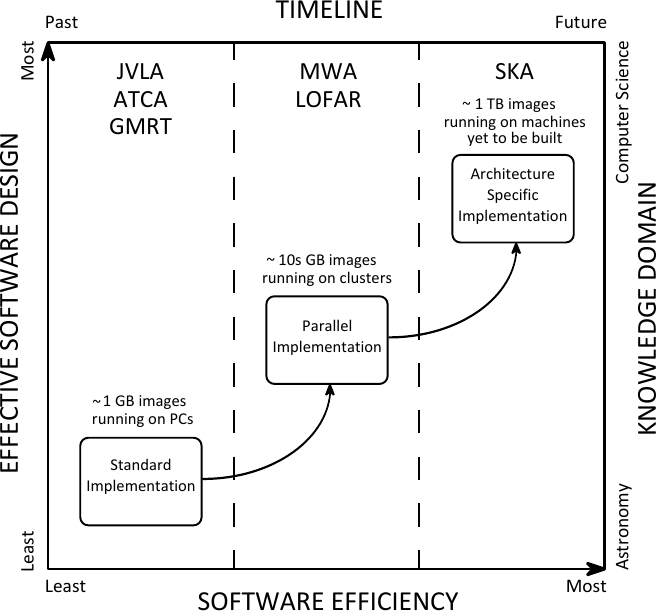}{P026_f2}{Schematic of the change in astronomical software requirements over time from PC-based solutions to architecture-specific implementation required for the SKA.}

While there has been a growing number of studies to examine the performance of point source detection software, these attempts have been primarily focused on the accuracy, completeness and reliability of the measurements, rather than the computational cost and efficiency (e.g., \citealp{hhn2012,hws2015}). In practice, the software which will be incorporated in the SKA pipelines must be accurate and reliable, and concurrently, scalable and effective in the SKA's big data regime. We have evaluated the computational performance of the two most commonly used point source detection packages employed in radio astronomy, i.e., AEGEAN and Duchamp.

\section{Scalability to SKA Levels}

In order to examine the computational performance of the AEGEAN and Duchamp packages, we measured their running time for analysing a series of simulated radio images and compared them with the time complexity of their underlying algorithms (Flood-fill and Lutz, respectively). The six artificial images were comprised of $(2^n \times 10^3) \times (2^n \times 10^3)$ pixels, where $n=1, 2, \dots, 6$, with a surface density of approximately one point source of random flux density per 4000 pixels.

Duchamp and Lutz algorithm processing times are shown in Figure~\ref{P026_f1}. Duchamp time complexity exhibits a significant deviation from linearity when processing larger (above $16,000 \times 16,000$ pixels) images, while the processing time of its core algorithm, Lutz, is linear. In case of AEGEAN, while the program represents a linear time complexity behaviour, we observed a low (less than 50\%) parallel efficiency in the parallelized fraction of the code, in other words, AEGEAN does not scale linearly with the number of processing cores. Our results suggest that while the underlying approaches scale with number of pixels present as well as the number of processors, the implementation themselves do not and thus are unsuitable for SKA-scale imaging.

SKA-level images can be expected up to the level of $80,000 \times 80,000$ pixels for a Nyquist sampled beam and full resolution. Details such as the suitability of memory access patterns to the SKA Science Data Processor (SDP) architecture will determine whether any of the extant implementations or a modified implementation will be used. However, our investigations of Duchamp and AEGEAN show that neither implementation follows the expected scaling to SKA-levels, in terms of processing time or parallelization efficiency, and both programmes cease to perform linearly, and therefore, are inadequate once images reach reasonable sizes (above $32,000 \times 32,000$).

This work has focused on point source detection which is a required task for both the local sky model and output science catalogues (see \citealp{O12.7_adassxxv}). However, there is a necessity to study the performance and functionality of diffuse and extended source detection software in the future, however, at present, this type of image analysis is in its infancy with only a very small number of algorithms capable of detecting low-surface-brightness diffuse and extended radio emission (e.g. \citealp{ffj2014,O2.4_adassxxv}). 

\section{Road to the SKA: a Collaborative Approach}

To date, efficiency of algorithmic implementation has not been the most pressing concern in astronomy as it has not been a limiting factor in information extraction. However, as we move from data sizes of less than 1 GB, which can reasonably be examined on a single pc, to images of 10s of GB needing compute clusters for generation and analysis, to the truly staggering levels of the SKA of TB images generated on a machine the likes of which does not yet exist, algorithmic efficiency becomes critical (see Figure~\ref{P026_f2}). As such we must move from a regime of astronomer generated algorithms and implementations to collaborative efforts by teams of astronomers and professional software engineers. The SKA will require a paradigm shift in which astronomers must develop the algorithms but then hand them over to software engineers for implementation. The astronomical community should therefore anticipate this shift and commence collaborations with software engineers now.

\bibliographystyle{asp2014}
\bibliography{P026}{}

\end{document}